\documentclass[aps,prl,twocolumn,superscriptaddress]{revtex4}
\usepackage[pctex32]{graphicx}

\begin{document}
\title{Phase engineering of squeezed states and controlled entangled number
states of Bose-Einstein condensates in multiple wells}

\author{Khan W. Mahmud}
\thanks{Present address: Department of Physics,
University of Michigan, Ann Arbor, MI 48109, USA}
\affiliation{Department of Physics, University of Washington,
Seattle, WA 98195-1560, USA}

\author{Mary Ann Leung}
\affiliation{Department of Chemistry, University of Washington,
Seattle, WA 98195-1700, USA}

\author{William P. Reinhardt}
\affiliation{Department of Physics, University of Washington,
Seattle, WA 98195-1560, USA} \affiliation{Department of Chemistry,
University of Washington, Seattle, WA 98195-1700, USA}

\begin{abstract}
We provide a scheme for the generation of entangled number states of Bose-Einstein condensates in multiple wells, and also provide a
novel method for the creation of squeezed states without severe
adiabatic constraints on barrier heights. The condensate ground state in a multiple well trap can be evolved, starting with specific initial phase difference between the neighboring wells, to a state with controllable entanglement. We propose a general formula for appropriate initial phase differences between the neighboring wells that is valid for any number of wells, even and odd.
\end{abstract}

\maketitle

Entanglement, a nonclassical correlation between two or more physical
systems, lies at the heart of the profound difference between quantum
mechanics and a local classical description of the world~\cite{epr1}. Apart
from their discussions in the philosophical and foundational aspects of
quantum mechanics~\cite{bell2}, entangled states in recent years have become
an essential resource for the emerging field of quantum information
processing. Entangled states have been created with photons~\cite{ghz2},
four atoms~\cite{fouratoms}, and most recently with many cold atoms in a Mott
insulating state in an optical lattice~\cite{mott3}. Cold atoms in optical
lattices have been a vibrant research area with several new observations such
as the superfluid to Mott insulator transition~\cite{mott1} and
number-squeezed states of Bose-Einstein condensate (BEC)~\cite{kasevich2}.
Entangled and squeezed states hold promise in studies related to quantum
measurement, the Heisenberg limited atom interferometry and quantum
computing and quantum communication protocols~\cite{huelga1}. While the
consequence of entanglement for an Einstein-Podolsky-Rosen (EPR) pair is
quantified in Bell's inequality ~\cite{bell1}, a more striking conflict
between quantum mechanics and local realism is exhibited by
three maximally entangled particles also known as the
Greenberger-Horne-Zeilinger (GHZ) states~\cite{ghz1}. GHZ state of $N$
particles has the form
\begin{equation}
|\Psi \rangle =\frac{1}{\sqrt{2}}\left( |1\rangle ^{\otimes
N}+|2\rangle ^{\otimes N}\right) \label{eqn:ghz}
\end{equation}
where 1 and 2 are the basis states for a two state (spin
$\frac{1}{2}$) system, and are written in the standard notation as
$|00...0\rangle+|11...1\rangle$.  In the occupation number basis,
$|1\rangle ^{\otimes N}$ and $|2\rangle ^{\otimes N}$ respectively
denote, $|N,0\rangle$ and $|0,N\rangle$. The two-state model has
been generalized to more than two spin components in
Ref.~\cite{cerf1}. The superposition of two macroscopically distinct
states, rather than simply the internal degrees of freedom, each
occupied by all $N$ particles, has been discussed by Schr\"{o}dinger
in the famous cat parable~\cite{schrodinger}; partial realization of
such states has been obtained with Josephson junction
loops~\cite{friedman1}.

In this paper, we discuss the generation of macroscopic entangled number
states of a multiwell BEC of the form
\begin{equation}
|\Psi \rangle =\frac{1}{\sqrt{M}}\left( |1\rangle ^{\otimes N}+|2\rangle
^{\otimes N}+...+|M\rangle ^{\otimes N}\right)
\label{eqn:multiwell}
\end{equation}
where $1,2,3,..M$ label the macroscopically and spatially distinct wells,
$|i\rangle ^{\otimes N}$ now denoting $|0,0...,N_{i}=N,...0\rangle$. We
show that states approximating the extreme entangled states of Eq.~(\ref{eqn:multiwell}) may be
generated in a controlled fashion by time evolution of
appropriately phase imprinted ground states of a multiwell BEC with periodic
boundary conditions for M=3 and 4. We show that the choice of initial
barrier heights, which determine the extent of ground state number
squeezing, and the rate of barrier ramping can be used to control
the extent of entanglement of the final states.  We also show that fully fragmented states can
be generated via natural time evolution from the ground state
following certain initial phase offsets. The creation of such
fragmented states through phase engineering and without the severe
adiabatic constraints on the rate of barrier height change provides
an alternative to the current experimental approaches~\cite{mott1,kasevich2}. Finally, based on results obtained for two, three, and four well configurations, we conjecture a generalized formula, for M wells, for the phase offset between neighborhing wells appropriate for the generation of number entangled states.

\begin{figure}[t]
\includegraphics[width=8.0cm,height=6.5cm]{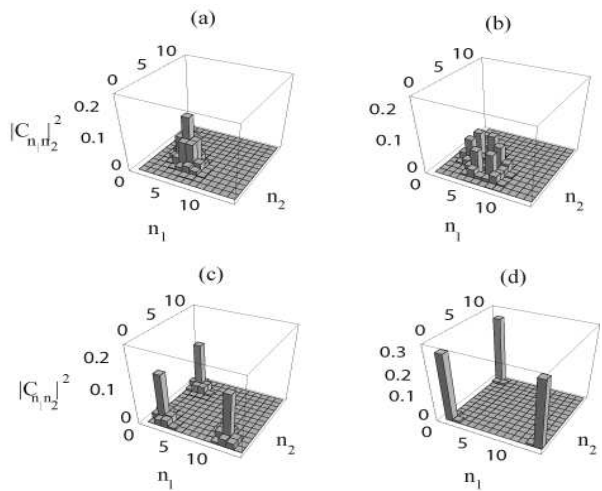}
\caption{Fock state coefficients for 12 particles in three wells: (a) the ground state, (b) 10th, (c) 76th and (d) 91st, the highest state. The ground state has a
Gaussian shape, while higher lying states are entangled number states. $n_1$ and $n_2$ are the Fock state indices and the vertical axis shows probabilities. Points beyond the cross-diagonal are unphysical.} \label{fig:stationary}
\end{figure}

We approximate the physics of a BEC in a multiwell potential by the
Bose-Hubbard model~\cite{bosehubbard1}. Thus
\begin{eqnarray}
\hat{H} &=&-J\sum_{i}(a_{i}^{\dagger }a_{i+1}+a_{i+1}^{\dagger
}a_{i})+\sum_{i}\epsilon _{i}\hat{n_{i}}  \nonumber \\
&&+\frac{1}{2}U\sum_{i}\hat{n_{i}}(\hat{n_{i}}-1)  \label{eqn:BHhamil}
\end{eqnarray}
where $\hat{n_{i}}=a_{i}^{\dagger }a_{i}$ is the number operator,
$J$ is the nearest neighbor tunneling term, $U$ is the on-site
energy, and $\epsilon _{i}$ is the energy offset of the $i$th
lattice. To simplify a theoretical study, we make a one parameter
approximation of the tunneling and mean-field terms:
$U/J=1/e^{-\alpha}$; and, for the symmetric wells explored here,
$\epsilon _{i}=0$. $\alpha $ is a dimensionless parameter that can be
mapped onto the barrier height. This parametrization allows a simple study of
continuous change of barrier height through the variation of a
single parameter $\alpha $. For example, for a lattice made of red
detuned laser with $\lambda =985$ nm and for $^{23}$Na, a barrier
height $15E_{R}$ gives $U=0.15E_{R}$ and
$J=0.07E_{R}$~\cite{bosehubbard1} where
$E_{R}=\frac{\hbar^{2}k^{2}}{2m}$ is the recoil energy from
absorption of a photon; these experimental parameters then
correspond to $\alpha =2.14$.

In order to gain insight into the types of stationary states possible for
the multiwell Bose-Hubbard model, we first analyze the quantum mechanical
properties of the simplest multiwell potential, $M=3$, assuming three
symmetric wells in a circular array~\cite{circularandlineararray}. The state vector is a superposition of
all the number states
\begin{equation}
|\Psi _{i}\rangle
=\sum_{n_{1},n_{2}=0}^{N}c_{n_{1},n_{2}}^{(i)}|n_{1},n_{2},n_{3}\rangle
\label{eqn:basis}
\end{equation}
Here $n_{1}$, $n_{2}$, and $n_{3}=N-n_{1}-n_{2}$ are the number of
particles in each of the three wells. Fig.~\ref{fig:stationary}
shows the Fock space probabilities,$\left|
c_{n_{1},n_{2}}^{(i)}\right| ^{2}$, for representative stationary
states for $N=12$ and $\alpha =0$ ($U/J=1$). Our method of graphical
representation is described in the figure caption. The ground state
in Fig.~\ref{fig:stationary}(a) is a broad Gaussian while the higher
lying states, Figs.~\ref{fig:stationary}(b)-(d), are number
entangled states of increasing {\it extremity} corresponding to
increasing numbers of particles simultaneously in all three wells,
the highest of which in panel (d) is an extreme superposition state
of the form $|N,0,0\rangle+|0,N,0\rangle+|0,0,N\rangle$. The number
of non vanishing Fock state coefficients determines {\it sharpness},
and thus (d) is sharper than (c).

\begin{figure}[t]
\includegraphics[width=6.5cm,height=8.2cm]{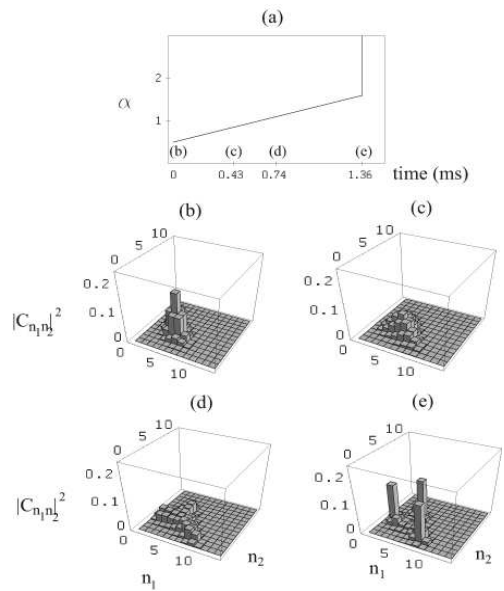}
\caption{Evolution to an entangled Fock space state: (a) barrier
ramp showing the location of the following time evolved states: (b)
initial state, (c) at 0.43 ms the Gaussian distribution broadens,
(d) at 0.74 ms the distribution is `splitting', (e) A three-peaked
state is formed at 1.36 ms; a macroscopic superposition of definite
number of particles simultaneously in all three wells.}
\label{fig:evolving}
\end{figure}
\begin{figure}[t]
\includegraphics[width=6.5cm,height=8.0cm]{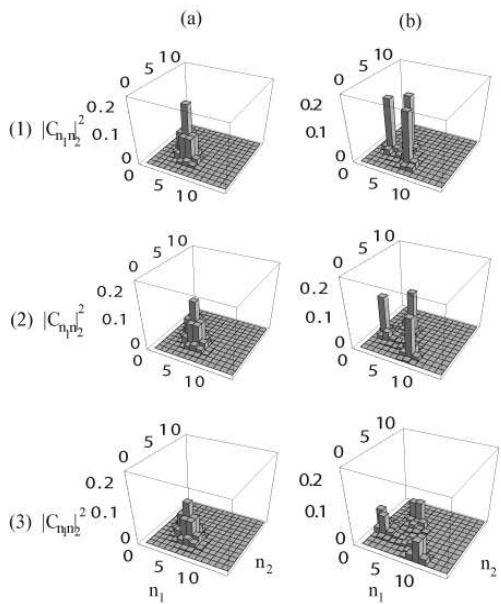}
\caption{Entangled states evolved from ground states with different
initial squeezing. Row (1) shows the states with $\alpha=1.5+t$: (a)
initial ground state and (b) final state. Row (2) is for
$\alpha=0.5+t$ and (3) is for $\alpha=t$. Column (b) gives the
states at t=1.85 ms, t=1.36 ms, and t=0.99 ms respectively. The
initial squeezing of the ground state thus determines the extremity
of the resulting entangled states.} \label{fig:tunable}
\end{figure}
\begin{figure}[t]
\includegraphics[width=7.2cm,height=7.2cm]{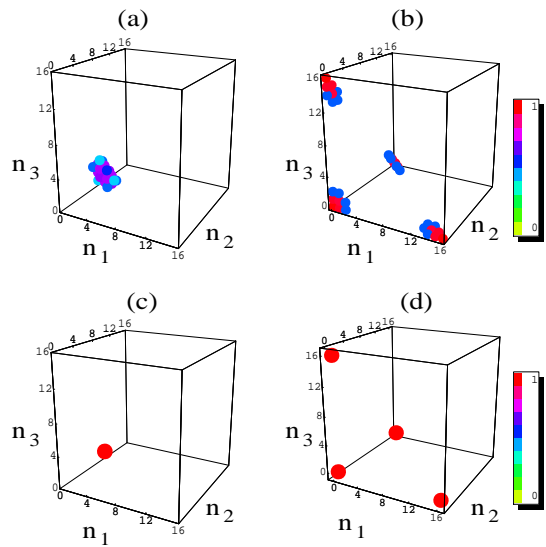}
\caption{Four well stationary and time evolved states:
(a) ground state, (b) highest excited state, (c) phase
engineered fragmented state following a $\pi/2$ relative phase shift of the ground state, (d) an entangled state evolved from the ground state following a $\pi$ relative phase shift. The three dimensions show the Fock state indices $n_1$, $n_2$ and $n_3$, probabilities are shown in the color intensity scale. For graphical clarity, only the points higher than $40\%$ of the highest probability are shown, with the highest probabilities normalized to 1.}
\label{fig:fourwell}
\end{figure}

It is unlikely that such maximally entangled states can be generated
via a sequence of single particle excitations. They may however, be
dynamically generated via phase engineering from the approriate
ground state. Writing phases on part of a condensate is
experimentally feasible via interaction with a far off-resonance
laser~\cite{billsoliton}, and is assumed to be sudden with respect
to the dynamics of the condensate~\cite{billsoliton}.
Mathematically, this corresponds to multiplying the coefficients in
an expansion of the type of Eq.(4) by $e^{in_{i}\theta _{i}}$, where
$|n_{1},n_{2},...n_{i},...\rangle$ is the corresponding Fock state,
and $\theta _{i}$ is the phase for particles in the $i$th well.
Entangled state generation, obtained via integration of the (linear)
time-dependent Schrodinger equation, is shown in
Fig.~\ref{fig:evolving}, following phase imprinting of an initial
phase difference of $\frac{2\pi}{3}$ between the neighboring wells,
and a simultaneous linear ramping of the barrier as $\alpha= t$, as
shown in Fig.~\ref{fig:evolving}(a) (t here is dimensionless).
Panels 3(b) shows the initial ground state; 3(c) at time 0.43 ms,
the distribution broadens; 3(d) at 0.74 ms, in the process of
splitting the state towards the three corners; and 3(e) at 1.36 ms a
sharp, although not extreme, entangled number state with its
signature of three major non vanishing expansion coefficients. The
times are given for a $^{87}$Rb condensate, $\lambda=840$ nm,
$a_{sc}=5.8$ nm, $J=0.04E_{R}e^{-\alpha}$ and taking  $U=0.04E_{R}$
as approximately constant for calculational purposes. When an
appropriately entangled state is reached the barrier is suddenly
raised to halt further evolution in n-space. For the parameter
values used here, a simple time evolution without any change of
barrier also produces an entangled state, however barrier ramping is
used here to sharpen the resulting state. Control of the extremity
of the states can be achieved by choice of the initial barrier
height, controlling the initial squeezing of the ground state. This
is demonstrated in Fig.~\ref{fig:tunable}, where different initial
squeezing have been used for rows (1), (2) and (3). The columns
show: (a) the ground state, and (b) the final state at the end of
the barrier ramping. It is important to be able to tune to less
extreme entangled states, as such states are more robust to loss and
decoherence~\cite{kmahmud3}. Phase imprinting with a phase
difference of $\frac{4\pi}{3}$ produces an equivalent state, with
different phase space dynamics.

The physics of the creation of these entangled states can be
understood in terms of the underlying classical phase space
dynamics. Entangled state generation in a double well has been
thoroughly analyzed in a semiclassical phase space
picture~\cite{kmahmud3}. In the semiclassical limit valid for large
$N$, the operators $\hat{a_{i}}$ can be approximated by the
c-numbers $\sqrt{ n_{i}}e^{i\theta _{i}}$, where $n_{i}$ and $\theta
_{i}$ are the number and phase of particles in the $i$th well. The
double well dynamics is then described by the Hamiltonian of a
nonrigid physical pendulum~\cite{smerzi1} with the number and phase
differences ($\Delta n$,$\Delta \theta $) between the wells as
conjugate variables. This system has two fixed points - (0,0) and
(0,$\pi$). The (0,0) is a stable equilibrium, while the (0,$\pi$) is
stable in the $\pi$-state regime ($UN/J<1$) and unstable otherwise
($UN/J>1$). Exploitation of the bifurcation characteristics of the unstable equilibrium generates entanglement. Taking the initial parameters such that there is an unstable hyperbolic fixed point, phase imprinting moves the ground state to the unstable point, and
the wave packet splits in the subsequent time evolution. Control of the barrier
height can then be used in three different ways to control the motion of
the wave packet and thereby the nature of the desired entanglement~\cite{kmahmud3}. First,
a simultaneous ramping of the barrier with the natural dynamics at the unstable fixed point has been empirically found to be useful in directing the desired evolution
of the wavepacket. Second, initial barrier height, that is the
initial squeezing, helps shape the initial wave packet stretching it into different regions
of accessible phase space; and, thirdly the initial barrier height sets the (negative) curvature of the potential at the hyperbolic fixed point, controlling the rate of splitting of the wave packet. All of these effects can be visualized for the double
well system in the appropriate phase space~\cite{kmahmud3}. Similar to the simple
double well, the triple well, $M=3$, can be thought of as two
coupled pendulums~\cite{penna1} with complicated dynamics. In the
($\Delta n$,$\Delta \theta $) representation, the unstable fixed
points are (0,0,$\frac{2k\pi }{3}$,$\frac{2k\pi }{3}$), $k=1,2$. All
the features of the double well entanglement generation apply to
the three well case, and thus, many of the insights from the two and three well dynamics can be extended to arbitrary number of wells in a circular
array.
%
%

Next, we explore the dynamics of the $\frac{\pi}{2}$ and $\pi $
phase shifting for the symmetric four well case.  First we look at
the ground and the highest excited state in
Figs.~\ref{fig:fourwell}(a) and (b) for $N=16$, $U=0.25$,
$J=e^{-\alpha}$, $\alpha=0.175$, and assuming the case of $^{87}$Rb
in the previous example. As expected, the ground state is an
approximate Gaussian and the highest state is an extreme entangled
number state. The fixed point dynamics for the $\frac{\pi}{2}$
configuration is such that, for a constant barrier height, the state
evolves into a number-squeezed state during its evolution towards an
entangled state. Fig.~\ref{fig:fourwell}(c) shows a fragmented state
(at 14 ms), with essentially exactly 4 particles in each well,
obtained by this phase engineering scheme. Due to the location of
the fixed points in the phase space, the entangled states generated
by the  $\frac{\pi}{2}$ imprinting are not as extreme or sharp as
those generated by the $\pi$ method; an example of the latter is
shown in Fig.~\ref{fig:fourwell}(d) (at 1.5 ms). We also found that
a $\frac{3\pi}{2}$ phase difference between the wells produced
entangled states that only differ by a symmetry from the
$\frac{\pi}{2}$ phase imprinted states. Fragmented states have been
observed in a 12 well optical lattice with adiabatic raising of the
barrier~\cite{kasevich2}. We have shown here that fragmented states
can also be created in a natural and efficient fashion without
adiabatic constraints. In comparing our results to the results of
Ref.~\cite {polkovnikov2}, we show that the $\pi$ configuration in
an even number of wells that they identified is a special case of
many phase imprint dynamics that generate interesting correlated
states in multiple wells. Their changes in system parameters is to
drive the system from stability to a regime of instability. On the
other hand, we take our system to be in the unstable regime and
demonstrate the controllability of entangled states with barrier
manipulation; potentially useful for experimental detection.

For the two, three, and four wells, we find $M-1$ distinct phase
differences between the neighboring wells for the multiwell fixed
points~\cite {penna1,penna2}. These are given by a general formula
$\frac{2\pi j}{M}$ where $j=1,2,....M-1$, with $M$ being the number
of wells, which gives a $\pi$ phase difference for the $M=2$ double
well, a $\frac{2\pi}{3}$ and $\frac{4\pi}{3}$ phase difference for
the $M=3$ triple well, and a $\frac{\pi}{2}$, $\pi$, and
$\frac{3\pi}{2}$ phase difference for the $M=4$ quadruple well
configuration - we demonstrated the dynamics generated by all of
these phase difference imprints. Note that the total change in phase
in the circular loop is a multiple of $2\pi $, a vortex like
condition. We thus propose a general formula for $M$ wells,
\begin{equation}
\Delta \theta =\frac{2\pi j}{M}\text{,}
\label{eqn:phase}
\end{equation}
for the constant phase offset between neighboring wells leading to
the dynamical generation of entangled states. Here $j=1,2,..,M-1$,
and Eq.~(\ref{eqn:phase}), being valid for any number of wells, even
or odd, provides a substantial generalization of the phase offset
mentioned in Ref.~\cite{polkovnikov2}, which is valid only for the
special cases of an even number of wells and for $j=M/2$ ($\pi$
phase offset). The multiplicity of Eq.~(\ref{eqn:phase}) is
prominent for large number of wells, e.g. for 12 wells, there are 11
phase offset possibilities. As in the three and four well case
considered, symmetries may prevent all the imprinting offsets of
Eq.(5) from generating independent dynamics.

In conclusion, we have demonstrated phase engineering schemes for
the generation of entangled number states and fragmented states of
BECs in multiple wells. By controlling the initial barrier height
and rate of ramping, the entanglement of the final state can be
tuned. We presented a novel series of formulae for the initial phase
difference between the neighboring wells that is valid for any
number of wells, even or odd, each having distinct properties. The
creation, characterization, and applications of
multidimensional/multipositional Schr\"{o}dinger cat states of atoms
remain largely unexplored experimentally, and the theoretical
ramifications of such states, should they be easily produced, are
just emerging.

This work was supported by NSF grant PHY-0140091 and DOE computational
science graduate fellowship program grant DE-FG02-97ER25308.


\end{document}